 \newlength\smallfigwidth
 \documentclass[aps,prb,twocolumn,floatfix,showpacs,amsmath,amssymb]{revtex4-1}

\usepackage{graphicx,graphics}
\usepackage{float}
\usepackage[hypertex]{hyperref}
\usepackage{subfig}
\begin{document}


\title{Vortex core dynamics induced by hole defects in antiferromagnetic nanodisks}

\author{R.\ L.\  Silva}
\email{ricardo.l.silva@ufes.br} 
\author{R.\ D.\ Pereira}
\email{rodrigo.dias@ufes.br}
\affiliation{ Departamento de Ci\^encias Matem\'aticas
e Naturais, Universidade Federal do Esp\'irito Santo, S\~ao Mateus, 29932-540,
Esp\'irito Santo, Brazil}
 
\author{A.\ R.\ Pereira}
\email{apereira@ufv.br} 
\author{W.\ A.\ Moura-Melo}
\email{winder@ufv.br} 
\author{L.\ A.\ S.\ M\'{o}l}
\email{lucasmol@ufv.br} 
\affiliation{ Departamento de F\'isica,
Universidade Federal de Vi\c cosa, Vi\c cosa, 36570-000, Minas
Gerais, Brazil }

\author{A.\ S.\ T.\ Pires}
\affiliation{ Departamento de F\'isica,
Universidade Federal de Minas Gerais, Belo Horizonte, 30123-970, Minas
Gerais, Brazil }

\date{\today}

\begin{abstract}
Direct observation of vortex states in an antiferromagnetic layer have been recently
reported [Wu, {\em et al}, Nature Phys. {\bf 7}, 303 (2011)]. In contrast to their analogues in ferromagnetic systems, namely in nanomagnets, the vortex core of antiferromagnets are not expected (and have not been observed) to present gyrotropic or any other remarkable dynamics, even when external fields are applied. Using simulated annealing and spin
dynamics techniques we have been able to describe a number of properties of such a vortex state. Besides of being in agreement with reported results, our results also indicate, whenever applied to antiferromagnetic nanodisks, that the presence of holes in the sample may induce two types of motions for this vortex. Its dynamics depends upon the relative separation between its core and the hole: when they are very apart the vortex core oscillates near the nanodisk center (its equilibrium position); while, if they are sufficiently close, the core moves towards the hole where it is captured and remains static.
\end{abstract}

\maketitle
\section{Introduction}
Ferromagnetic (FM) disks of micron and submicron size are characterized by an in-plane close flux magnetization, minimizing the dipolar energy. However, to minimizing the exchange energy at the disk center, the magnetization revolves against the plane, in a small area of only a few exchange lengths, creating the vortex core with two distinct polarizations, {\em up} ($p=+1$) or {\em down} ($p=-1$). Along with the vortex chirality, which accounts for the in-plane rotation sense, ($q=+1$) $q=-1$ for (counter-)clockwise, polarization are shown to govern many fundamental properties of vortex-state in FM nanodisks, such as its gyrotropic mode \cite{Choe04}, switching of the vortex core (polarity reversal)\cite{Van06,Yamada07,Hertel07} and vortex-pair excitation. A number of technological applications has been proposed as long as we may control one (or both) of these parameters on demand.\\

The scenario is expected to be quite distinct in antiferromagnetic (AFM) nanodisks where the vortex appears as a composed pattern of two anti-aligned sub-lattices, presumably yielding a much more rigid structure with no global dynamics, as those observed for its ferromagnetic counterparts. Actually, indirect evidence for this state have been obtained by inducing FM-ordered spins in AFM disks \cite{Sort06,Tanase09,Salazar09}. Only very recently, direct observation of imprinted AFM vortex states in $CoO/Fe/Ag(001)$ disks (an AFM layer of an AFM/FM bilayer system) has been reported\cite{Wu11}. In these systems, there appears a FM-AFM exchange coupling, in such a way that by measuring certain FM spin properties it provides a probe for their analogues throughout the antiferromagnet, like the spin-flipping field, the surface order parameter, the crystalline anisotropy, the domain wall width etc. To our knowledge, a systematic, experimental or theoretical, analysis concerning vortex dynamics in AFM nanodisks is still lacking, although a number of works has been devoted to study other aspects of such excitations in general AFM frameworks\cite{Wysin1991,Bogdanov1998,Chetkin2001,Guslienko2007}. From the theoretical point of view, the
vortex dynamic in layered antiferromagnets is very different whenever compared to ferromagnetic materials. For instance, these systems are ``Lorentz invariant'' and they bear no net magnetization. It is also expected that applied magnetic fields causes no vortex motion in AFM nanodisks, in deep contrast to their ferromagnetic analogues, a fact brought about by our present results. On the other hand, we argue that litographically inserted defects may introduce a new pathway for magnetization dynamics, including vortex core dynamics, in AFM nanosystems. This is a very interesting topic by itself, capable of feeding further investigation and potential applications of these AFM systems, similarly to what has occured to ferromagnetic disks in presence of cavities and other defects
\cite{Rahm03,Rahm04,Rahm+04,Silva08,Silva09,Afranio2005,Antonio2007,Winder2008,Cap-livro2009,Uhlig,Kupper,Compton,Hoffmann}. In addition, we also show that two types of vortex core dynamics takes place in
AFM disks as induced by a hole. Explicitly, our results show that if the vortex core and the defect are quite apart, then the core oscillates around its equilibrium position (a point near to the disk center), while if they are sufficiently close then the core moves towards the cavity where it is captured and eventually trapped. In both cases such a vortex core dynamics appears as a result of the competition between the two (AFM) sub-lattices to minimize exchange energy in the presence of the hole, inside which exchange cost vanishes.\\
\begin{figure}[hbt]
\includegraphics[angle=0.0,width=\columnwidth]{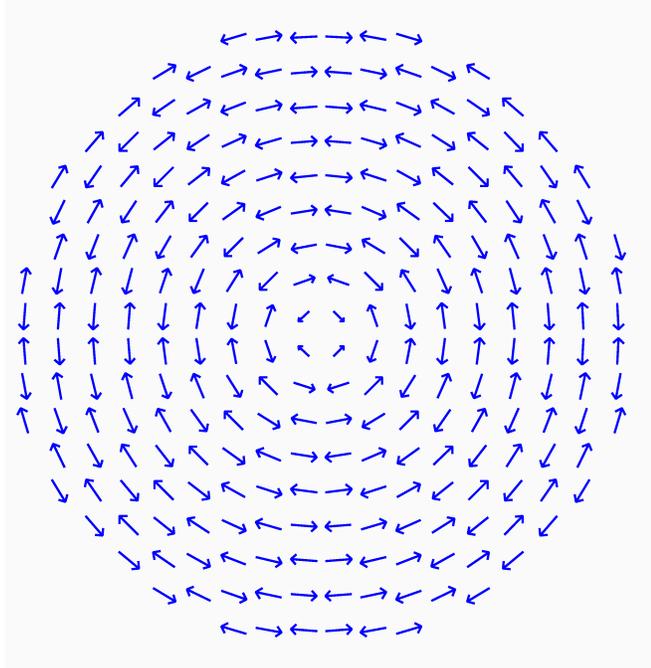}
\caption{ \label{ant1}  Top view of an antiferromagnetic nanodisk with a vortex state. Namely, note the two anti-aligned sublattices.}
\end{figure}

\begin{figure}[hbt]
\includegraphics[angle=0.0,width=\columnwidth]{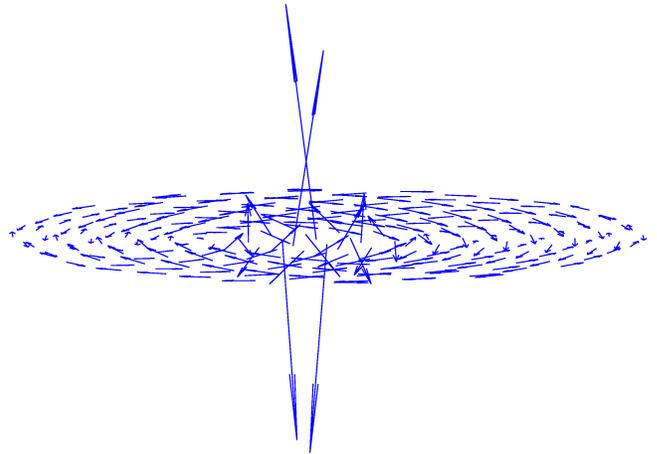}
\caption{ \label{ant2} Lateral view of the vortex configuration from Fig. \ref{ant1}. In the vortex core, the central region, dipoles develop out-of-plane projections. It must be emphasized that a AFM vortex consists of two {\em subvortices}, each one with its own polarity and chirality. In this case, one has $(p,q)=(+1,-1)$ and $(-1,+1)$, so that as a whole, no net $p$ neither $q$ shows up.}
\end{figure}

Before starting our analysis, it is useful to describe how a AFM-vortex can arise in
nanodisks. The vortex state in finite magnetic samples may emerge as the ground-state resulting from the competition between exchange and
magnetostatic interactions (if we focus on isotropic materials, anisotropic costs may be neglected). In these systems, such as magnetic disks, densities of effective magnetic charges are distributed in their volumes and surfaces, given by $\rho_M=-\vec{\bigtriangledown}\cdot \vec{M}$ and $\sigma_M=\vec{M}\cdot \hat{n}$, where
$\hat{n}$ is the unit vector normal to the surface at every point, while $\vec{M}$ accounts for the magnetization. Therefore, to minimize the magnetostatic energy the magnetization tends to point parallel to the dot surface. However,
as one goes towards the center, exchange energy density increases enormously (in a continuum description, it is scaled with $r^{-2}$, with $r$ being the distance from the disk center). Then, to regularize exchange
cost, magnetic moments tend to resolve against the plane, developing out-of-plane
projection, so that exactly at the center it is perpendicular to the disk face. Such a behavior is followed by each sublattice in a fifnite AFM system. Finally, it must be remarked that the AFM vortex state always appears as a composite pattern with opposite polarity and chirality at the sublattices, so that both parameters identically vanish for the whole vortex (see Figs. \ref{ant1} and \ref{ant2}; further details below). This is the reason why its gyrotropic factor also falls off preventing dynamical manifestations, as the translational gyrotropic oscillation of the vortex core observed in ferromagnetic
nanodisks\cite{Choe04,Van06,Yamada07,vortexcore1,vortexcore2}.\\

\section{Methods and Results}
In order to study such a state in very thin AFM disks, with thickness
$L$ and radius $R$, so that the aspect ratio $L/R\ll 1$, we firstly focus on each sublattice and assume that
$\vec{\bigtriangledown}\cdot \vec{M}\approx0$ there (this is quite reasonable, except perhaps in a very small region very close to the vortex center). Thus, the only source of magnetostatic interactions
could come from superficial magnetic charges, $\sigma_M$. Eventually, the vanishing of net magnetization in antiferromagnets implies that magnetostatic energy also falls off whenever computed throughout the whole sample, preventing stability of domains in AFM systems, once typical gain in configurational entropy is not sufficient to overcome domain wall energies.
Now, to describe the system we pursue an alternative strategy substituting the AFM/FM bi-layer sample by a two-dimensional film containing an effective magnetostatic anisotropy coming from both FM and AFM
exchange couplings (see Refs. \cite{Silva08,Silva09} for applications to FM nanodisks, even in presence of defects). Now, the film is embeded with a regular square lattice defined inside a circumference of radius $R$ where magnetization vector is distributed over the sites. In addition, the magnetostatic interactions due to the presence of the magnetic charges in the lateral and top surfaces of the film are replaced by local potentials. The model is summarized by the following Hamiltonian:
\begin{equation}\label{eq1}
H_{0} = -\sum_{i,j}J_{i,j}\vec{\mu}_{i}\cdot \vec{\mu}_{j}
+\sum_{\alpha=1,2}\sum_{k}\lambda_{\alpha}(\vec\mu_{k}\cdot\hat{n}_{k,\alpha})^{2} +\sum_{i}\vec{h}
\cdot \vec{\mu}_{i},
\end{equation}
where $J_{ij}=0$ for sites inside the hole and $J_{ij}<0$ for remaining sites of the film. Here
$\mu_{i}=\vec{M}_{i}(\vec{r})/M_{s}=\mu_{i}^{x}\hat{x}+\mu_{i}^{y}\hat{y}+\mu_{i}^{z}\hat{z}$ is
the unit spin vector at position $i$ ($M_{s}$ is the saturation magnetization) and the sum ${i,j}$
is performed over the nearest-neighbors spins. The term with positive constant, $\lambda_{\alpha}$, mimics the
magnetostatic energies at the top face of the disk, $\lambda_{1}$, while $\lambda_{2}$ accounts for the
lateral edge. In turn, $\vec{h}$ is the external magnetic field. The hole defects are introduced by
removing from the system a number of neighbor spins around a site $\vec{r}_{i}$. Of course, the number of neighbors spins removed around a particular position
($\vec{r}_{i}$) defines the hole size $\varrho_{i}$. The local unit vectors $\hat{n}$ must be normal to the surfaces at every point, so that $\hat{n}_{k,1}=\hat{z}$ for the disk face (in the $xy$-plane), while
$\hat{n}_{k,2}=\vec{R}/|\vec{R}|=\hat{R}$along the disk circumference of radius $R$.\\
\begin{figure}[hbt]
\centering
\subfloat[Initial position of vortex core in the center of disk]
{\label{figb1:start}\includegraphics[angle=0.0,width=\columnwidth]{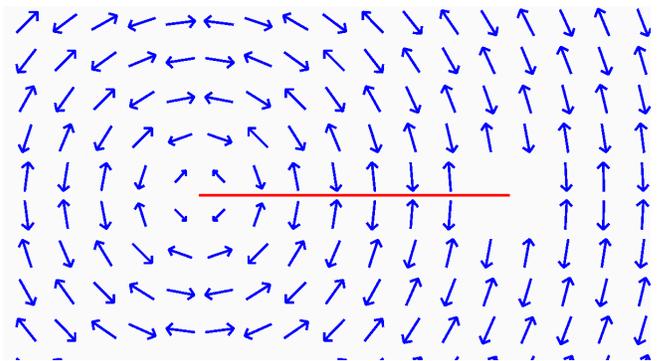}}\\
\subfloat[Vortex core in a position between the disk and hole centers]
{\label{figb2:meio}\includegraphics[angle=0.0,width=\columnwidth]{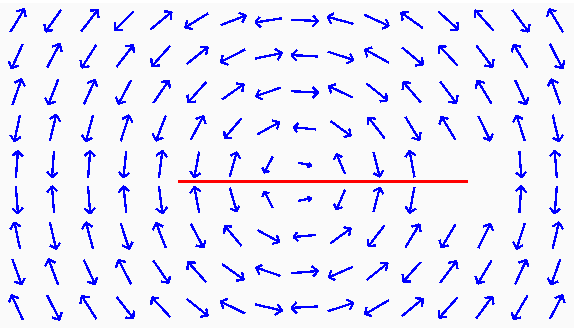}}\\
\subfloat[ The final position of vortex core (vortex trapped by the hole)
]{\label{figb3:fim}\includegraphics[angle=0.0,width=\columnwidth]{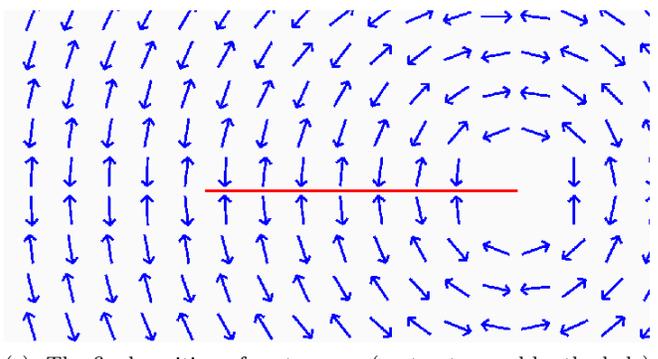}}\\
\caption{ \label{ant3}  Partial view, showing only a part of the disk, considering the vortex core
and the hole. The sequences of the pictures show the movement of the vortex in direction to the
hole.}
\end{figure}

In our calculations we have used simulated annealing and spin
dynamics simulations for square lattices occupying all possible points inside a circumference with
radius $R$ (in most simulations we have used $R=10a$, $R=20a$, $R=25a$ and $R=40a$; $a$ is the lattice spacing constant). It
is solved by employing the simulated annealing and the fourth-order predictor-corrector method. The results below have taken $\lambda_{1}=0.2$ and $\lambda_{2}=2.0$. Other values for $\lambda_{1}$ does not alter the essential physics reported here provided that $0<\lambda_{1}<0.28$; for $\lambda_{1}>0.28$, the
vortex become essentially planar and does not develop out-of-plane components (similar scenario occurs for $1.8<\lambda_{2}<2.2$). The vortex structure itself in an AFM nanodisk was obtained by using eq.(\ref{eq1}) and the simulated annealing techniques. The vortex configuration so emerged may be viewed as a composite structure of a pair of {\em sub-vortices} each of them associated with one sublattice. Each {\em sub-vortex} has a pattern quite similar to its counterpart in FM disks, but always appear with opposite polarity and chirality each other, yielding net vanishing of such quantities for the composite vortex (see Figs. \ref{ant1} and \ref{ant2}). After being formed and stabilized, spin dynamics techniques are used to study how the composite vortex behaves when constant or sinusoidal, $\vec{h}_{ext}=\vec{h}_{0}\sin(\omega t)$, external magnetic fields parallel or perpendicular to the plane of the disc are applied: in
contrast to what happens to a vortex in FM disk, here its core remains static at the disk center (its equilibrium position), despite the intensity, frequency, and direction of the field, in agreement with theoretical expectations\cite{****} and experimental observation\cite{*****}.\\

Dynamics has been turned on whenever defects are introduced, by removing some spins around a given site. Now, the hole induces translational motion of the vortex core yielding two very distinct consequences: an oscillatory motion of the vortex core near the disk center whenever the cavity is far away the core; or, it may be captured if placed sufficiently close to the hole. To be more precise, we have inserted a single hole of radius $\varrho=a$ (with four neighbor spins removed) in the system. We have chosen the origin of the coordinates to be the vortex center, which coincides with disk center, $(x_0,y_0)=(20a, 20a)$ (for a lattice composed by $40\,{\rm x}\,40=1600$ spins), and the simulations have considered the presence of a unique hole, introduced somewhere else. Explicitly, we have taken the hole center to be located at ten different points along the $x$-axis, say, at $\vec{r}=(a,0),(2a,0),\ldots (10a,0)$. Our results have shown that for $a\leq r \leq 8a $, the vortex core moves towards the hole, where it is captured yielding a practically static coreless vortex state (see Fig. \ref{ant3}). On the other hand, whenever the inserted hole is sufficiently apart from the vortex center, $r>=9a$, its core develops an oscillatory motion around a position very close to the disk center (along the line joining the hole center and disk center), whose amplitude depends on the distance core-hole, diminishing as $r$ increases. We should remark that intentionally inserted holes in FM disks have attracted a great deal of efforts in the last years and a  number of effects on the vortex state has been
predicted\cite{Silva08,Silva09,Afranio2005,Antonio2007,Winder2008,Cap-livro2009,Felipe2010} and experimentally
confirmed\cite{Rahm03,Rahm04,Rahm+04,Uhlig,Kupper,Compton,Hoffmann}.\\ 

\section{Conclusions}
In this paper we have studied finite size circular antiferromagnetic thin films. The composite vortex state formed by the two AFM sublattices is show to be dynamically insensitive to any applied field. However, dynamics may be induced by hole defects, where vortex exchange energy is minimized. Once AFM vortex has no net gyrotropic vector its associated motion cannot show up (contrasting with defect induced gyrodynamics in FM vortex, Ref.\cite{Silva08}); rather, its subsequent translation gives rise to oscillations around its equilibrium position, if the vortex core and defect are very apart, while if they are placed close, the hole captures the core, yielding to a static corelless vortex as the final state. Our results could feed further investigation on AFM nanoscaled samples with inserted cavities, mainly when such induced vortex dynamics concerns.

\begin{acknowledgments}
The authors thank CNPq, FAPEMIG and CAPES (Brazilian agencies) for financial support.
\end{acknowledgments}

\end{document}